\documentstyle[epsf,preprint,aps,amsfonts,amssymb]{revtex}
\tightenlines
\begin{document}

\draft

\title{ The structure relaxation of carbon nanotube}
\author{ Xin Zhou$^{1}$, Hu Chen$^{2}$, Jianjun Zhou$^{1}$ and 
Ou-Yang Zhong-can$^{1,2}$}
\address{$^{1}$Institute of Theoretical Physics, Academia Sinica, P. O. Box
2735, Beijing 100080, China
\\ $^{2}$Center for Advanced Study, Tsinghua University, Beijing 100084,
 China}
\date{\today}

\maketitle

\begin{abstract}
 A simple macroscopic continuum elasticity theory (CET) is
 used to calculate the structure relaxation of single-wall 
 carbon nanotube (SWNT), an analytic formula is obtained. 
We also expand an atomic scale three-parameter empirical model 
[ T. Lenosky {\emph et al.} Nature 355, 333(1992)] in order to 
correctly describe the bond-length change effects. The structure
 relaxation of SWNT expected by the model is good in agreement
 with our CET results, and very well consistent with the previous
calculation from a first principles local density function 
approximation. Using the expanded Lenosky model, we calculate the strain 
energy of bending tube. The obtained results are good in agreement
with the previous theoretical expectation.  
It shows the model may be a good simple replacement of
 some more sophisticated methods on determining carbon networks 
 deformations.
\end{abstract}



\newpage

Recently, carbon nanotubes with cylindrical graphite structures 
have been intensively
 investigated~\cite{Dekker}, and many experimental and theoretical 
 researches have been 
 performed~\cite{Hamada,White,Wildoer,Dresselhaus}.  
The structure of nanotubes is qualitatively well 
known through the simple construction of rolling 
a perfect graphene sheet, where only one parameter is to be 
determined: The lattice parameter or a bond-length. The 
first-principles local density function 
approximation (LDA)~\cite{Blase,Daniel}, 
tight-binding (TB) approximation~\cite {Hamada} and empirical 
potential methods 
~\cite{Tersoff,Yakobson,Lu,Hernandez} have 
been used to determine the bond-length of tube.
In all these theoretical studies, a small relaxation effect 
has been found, the bond length of tubes is not equal to that 
of graphene sheet since the curvature of the tubes and thus 
the structural inequivalence between the axis
and normal direction render the carbon-atom hexagons distorted, 
although the details of 
the bond relaxation from different authors are different. In this 
letter, we use a simple 
macroscopic continuum elasticity theory (CET) and an expand 
microscopic empirical model to calculate the bond 
length relaxation of single-wall carbon nanotubes (SWNTs).
We find from both the two methods: 
(1) the radius of any SWNT is slightly larger 
than the expected value from the rolling graphene sheet. 
(2) The radius relative increasing of SWNT is reverse 
proportion to the square of the tube radius, and independent or 
slightly dependent on the helicity of tubes. 
Our results are in good in agreement with the previous LDA 
results. 
(3)Although the energy contribution of 
bond-length change is very small in straight SWNT~\cite{Lenosky}, 
it can strongly affect the structure relaxation of tubes.
(4) We use the expand model to describe the bending of nanotube, and
 find the obtained results are good consistent with the previous 
 theoretical expectation.
  
SWNT can be indexed by a pair of integers ($n_1$, $n_2$)~\cite {White}. 
From the rolling graphene model, we have
\begin {eqnarray}
   \rho_0=\frac{{\sqrt{(n_1^2+n_1 \cdot n_2 +n_2^2)}} \cdot a}{2 \pi},
\end {eqnarray} 
where $\rho_0$ is the non-relaxed radius of SWNT, $a=\sqrt{3} r_0$ is the 
lattice constant of graphite sheet,
$r_0=1.42$ $\AA$ is the bond length of graphene. Supposing the 
relaxed radius is $\rho$, we define 
\begin{eqnarray}
    {\epsilon}=\frac{\rho-\rho_0}{\rho_0}.
\end{eqnarray}

A number of theoretical and experimental studies show the deformations
of graphenes and SWNTs can be well described by the 
CET~\cite{Daniel,Yakobson,Lu,Ouyang,Zhou1}. The rolling    
 energy of graphene is only dependent on the rolling 
 radius, and the stretching energy is only dependent on the relative
 elongation, independent on 
the rolling and stretching direction of graphite sheet, and we have 
\begin{eqnarray}
  E_r=C/\rho^2 ,  \\
  E_s=\frac{1}{2} E^{''} \cdot {\epsilon}^2. 
\end{eqnarray}
 
Here $E_r$ and $E_s$ are the rolling and stretching energy per atom, 
respectively. 
$C \approx 1.2 \sim 2.0$ $eV {\AA}^2$, depending on the different models, 
$E^{''} \approx 58 \sim 59$ $eV$
~\cite{Daniel,Yakobson,Lu,Ouyang,Zhou1}. 
When rolling a graphite sheet to SWNT, the bonds are stretched in 
the circumference direction of tube, ( we neglect the possible
 smaller strains in the axis direction). The strain energy per atom 
is the sums of the rolling and stretching energy. 
\begin {eqnarray}
 E_t=\frac{1}{2} E^{''} \cdot {\epsilon}^2 +\frac{C}{{\rho_0}^2 (1+{\epsilon})^{2}}
\end {eqnarray}
Minimizing the strain energy, we obtain 
\begin {eqnarray}
  {\epsilon} \approx \frac{2 C}{E^{''}+6 C/{\rho_0}^2} \frac{1}{\rho_0^2} 
    \approx \frac{2 C}{E^{''}} \frac{1}{{\rho_0}^2}
\end {eqnarray}
So we obtain, ${\epsilon} \approx \frac{0.05}{\rho_0^2}$, 
(selecting $C=1.44$ $eV{\AA}^2$~\cite{Zhou1} )where the unit of $\rho_0$ is $\AA$. 
The results show: (1) any tube is relaxed when it is rolled from 
graphite sheet and the relaxation is slight but obvious;  
It is consistent with the previous theoretical expection~\cite{Hamada,Blase,Daniel,Kurti} 
 (2) Although the stretching energy from the 
 dilation of tube radius $E_s$
 is far smaller than the curvature energy contribution $E_r$, can be safely 
 neglected~\cite{Lenosky,Zhou1}, the size of the radius dilation 
 $\epsilon$ is sensitively dependent on $E_s$.
  
For more carefully discuss the relaxation effects, we also consider 
a microscopic model.
Lenosky {\emph et al.} employed a three-parameter empirical model 
which can well describe
the structure of the curved carbon networks while the bond lengths 
are constrained to $r_0$. 
Ou-Yang {\emph et al.}~\cite{Ouyang} reduced the model to a 
two-elastic-modulus macroscopic continuous
model and well treated the deformed energy of multi-wall carbon 
nanotube. However, for describing some deformations which the bond
 lengths are obvious changed, such as bending or stretching tube, 
 we need a more complete model which must including the contribution 
 from the bond length~\cite{Lenosky},  
\begin {eqnarray}
 &&E_t={\cal E}_0 \sum_{(ij)} \frac{1}{2} (r_{ij}-r_0)^2 + 
{\cal E}_1 \sum_{i} (\sum_{(j)} {\hat u}_{ij} )^2 \nonumber \\
&&+{\cal E}_2 \sum_{(ij)} (1- {\hat n}_i \cdot {\hat n}_j)
+{\cal E}_3 \sum_{(ij)} ({\hat n}_i \cdot {\hat u}_{ij})({\hat n}_j \cdot {\hat u}_{ij})
\label{eq7}
\end {eqnarray}
where ${\hat u}_{ij}$ is a unit vector pointing from carbon atom $i$ 
to its neighbor $j$, 
$r_{ij}$ is the distance between the atom $i$ and atom $j$, and 
${\hat n}_i$ is a unit vector normal to the fullerene surface at 
atom $i$. The summation $\sum_{(j)}$ is taken over the three 
nearest neighbor $j$ atoms on atom $i$, 
and the summation $\sum_{(ij)}$ is taken over only the nearest 
neighbor atom pairs. Supposing $r_{ij}=r_0$, fitting to 
LDA calculation, Lenosky obtained the values of (${\cal E}_1$, 
${\cal E}_2$, ${\cal E}_3$) which are $(0.96, 1.29, 0.05)$ $eV$, 
respectively~\cite{Lenosky}.

Using an empirical 4th-neighbor force-constant model of 
graphite~\cite{Rajishi}, which has been 
well described the different deformation of both graphite and 
SWNT~\cite{Lu} and correctly calculated the elastic modulus, 
we can compute the strain energy of SWNTs
 with different deformations. Fitting the values of $E^{''}$ 
 and Poisson ratio $\sigma$ 
of SWNT and graphite sheet which are calculated  
 from the force-constant model, we obtain   
 the value of ${\cal E}_0$ is $57$ $eV/{\AA}^2$. 
Since the bond-length change of SWNT is determined by the first 
order derivative of energy rather than energy itself, 
although the first term energy of eq. (\ref{eq7}) 
 in straight SWNT is very small~\cite{Lenosky,Zhou1}, 
the value of ${\cal E}_0$ still strongly affect the equilibrium 
structure of SWNT. 

The carbon atom positions of SWNT can be located from the rolling 
graphene sheet~\cite{Zhou1}, 
the vectors ${\vec u}_{ij}$ can be easily obtained, 
\begin {eqnarray}
 {\vec u}_{ij}=(r_0 {\sin}{\theta_{ij}}) {\hat e}_z
 - 2 \rho_0 {\sin}^2 {\frac{{\phi}_{ij}}{2}} {\hat e}_{r_i}
 +\rho_0 {\sin} {\phi}_{ij} {\hat e}_{{\phi}_i},
\end {eqnarray}
where $\theta_{ij}$ is the angle between ${\vec u}_{ij}$ and 
the circumference direction.
${\hat e}_z$, ${\hat e}_{r_i}$ and ${\hat e}_{{\phi}_i}$ are 
the unit vectors of 
axis, radial and tangent direction at atom $i$, respectively. 
${\phi}_{ij}$ is the relative 
azimuth between the atom $i$ and the nearest neighbor atom $j$,
independent on the site $i$, we have 
\begin {eqnarray}
 {\phi}_{ij}=\frac {r_0 {\cos} {\theta_{ij}}}{\rho_0}.
\end {eqnarray} 
After considering the relaxation effect, the radius and bond 
length will be stretched.  
 ${\vec u}_{ij}$ is approximately obtained, if $\rho_0$  
is replaced by the relaxed radius $\rho$, and ${\phi}_{ij}$ is not 
changed, where we have neglected the higher order terms. 
Then the total strain energy $E_t$ can be calculated. We can obtain
the analytic formula of $E_t$, then minimizing $E_t$ calculate 
$\epsilon$, but the obtained formula is very complex.
From the numerical calculating, 
we find the same square reverse proportion relation between 
$\epsilon$ and $\rho_0$ which is expected by CET. 
The coefficient is slightly dependence on the helicity of tube, 
since it only depends on an  
average of ${\cos}^{n} {\theta}_{ij}$ in three bonds. 
Fig. 1 shows the radius dilation of some SWNTs expected by 
the expanded Lenosky model. It is 
 very good in agreement with the characteristics $\rho_0$ square reverse 
proportion relation. As comparison, a first principles
LDA calculation also be depicted in Fig. 1, which is well 
consistent with our results. 
Since the structure relaxation can be obviously affected by 
the energy term from bond length change, our obtained value of 
$\cal E_0$ from the relaxation is more exact than from the strain 
energy itself of tube. Therefore,  
our result shows the expanded Lenosky model 
which ${\cal E}_0$ is fitted from the force constant model of graphite
 can be well used in the studying of the carbon network deformations.

In our previous paper~\cite{Zhou1} using Tight-Binding (TB) 
approximation,
we have successly calculated the strain energy of the bending tube, 
and simultaneously obtain the Young's modulus and the effective 
wall thickness of SWNT. It clarified the discrepancy on the wall
thickness of single-layer structure~\cite{Yakobson,Lu,Hernandez}. 
The expanded Lenosky model 
can be used to study the same problem, too.
Fig. 2 shows the strain
 energy per atom $E_b$ of the (5,5) tube as a function of the bending
 radius $R^2$. The data follow quite well the expected behavior 
 $E_b=E_r+\lambda/R^2$, where $E_r$ is the rolling energy of tube. The
 fitting value of $\lambda$ is about $185$ ${\rm eV {\AA}^2/atom}$, 
 slightly larger than the TB result $173$ ${\rm eV {\AA}^2/atom}$~\cite{Zhou1}
Fig. 3 shows $\lambda$ is a linear function of $\rho^2$, 
$a^{*}+b^{*} \ \rho^2$, which 
is consistent with the relation
expected by CET, $\lambda=\Omega \ Y \ b \ (\rho^2+b^2/4)$,
Where $\Omega=2.62$ ${\rm {\AA}^2/atom}$ is the occupied
 area per carbon atom in SWNTs, Y is Young's modulus, b is the wall 
 effective thickness, and $\rho$ is the radius of tube. The value 
 of $b^{*}$ is $16.2$ ${\rm eV/atom}$, slightly larger than the TB 
 results $15.3$. Since $\lambda$ is far larger than $a^{*}$, it is
 difficult to exactly fit the value of $a^{*}$ from the calculated
 data, due to the calculating errors. $a^{*} \approx 0.4$, if only 
 using the data of $(n,n)$ and $(n,0)$ tube, but $a^{*} \approx 1.5$, 
 if using the total data, where the unit of $a^{*}$ is 
 ${r_0}^2 {\rm eV {\AA}^2/atom}$ . However, our obtained the value of $b^{*}$
 is insensistive to the selected data, and $a^{*}$ is  
 between $0.4$ and $1.8$, 
 consist with the TB result $1.05$~\cite{Zhou1}.  
 Fig. 4 shows the strain energy of straight tube 
 ($1/R=0$) $E_r$ is reverse proportion to the square of tube 
 radius $\rho$, 
 the coefficient is about $1.48$ ${\rm eV {\AA}^2/atom}$,  
 good consistent with the TB calculation $1.44$ ${\rm eV {\AA}^2/atom}$. 
In previous TB approximation, since the unit cell of chiral tube
is very large, calculating strain energy is an arduous work, and
the method include twelve parameters. We think the expanded 
Lenosky model can well be used to determine the carbon network
structure.

In summary, our studies show the tube relaxation is inverse 
proportion to the square of the tube radius. The expanded Lenosky
model which only include four parameters can well describe the 
deformation of the bending carbon nanotube. It may be a simple 
replacement of some complex methods in determining the equilibrium
structure and the deformed energy of carbon networks.
  
The authors acknowledge the discussion with Dr. Haijun Zhou.

\begin{figure}
\centerline{\epsfxsize=10cm \epsfbox{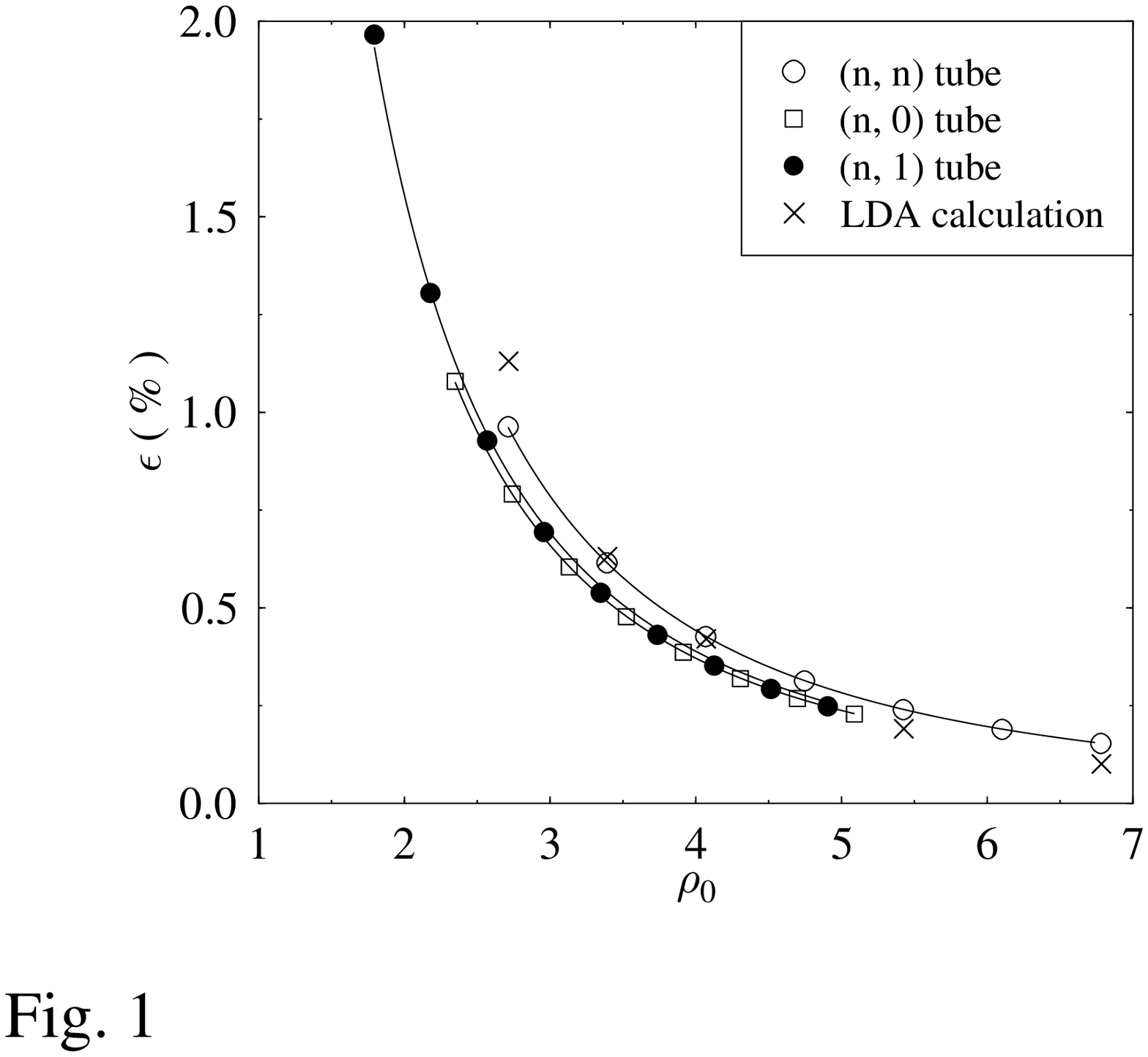}}
\caption {
The radius dilation of SWNT $\epsilon$ versus the radius of 
tubes. The solid line drawn across the
data corresponds to a least square fit to the $\frac{\alpha}{{\rho_0}^2}$ 
behaviors. The $\alpha$ values of the $(n, n)$ $(n, 0)$ and $(n, 1)$ tubes are
0.071, 0.059 and 0.062 ${\AA}^2$, respectively, slightly depend on the 
helicity of SWNT.  
The results of LAD is
from the fig. 1 (c) of ref. {\protect\cite{Daniel}}.
\label{fig1}
}
\end{figure}

\begin{figure}
\centerline{\epsfxsize=10cm \epsfbox{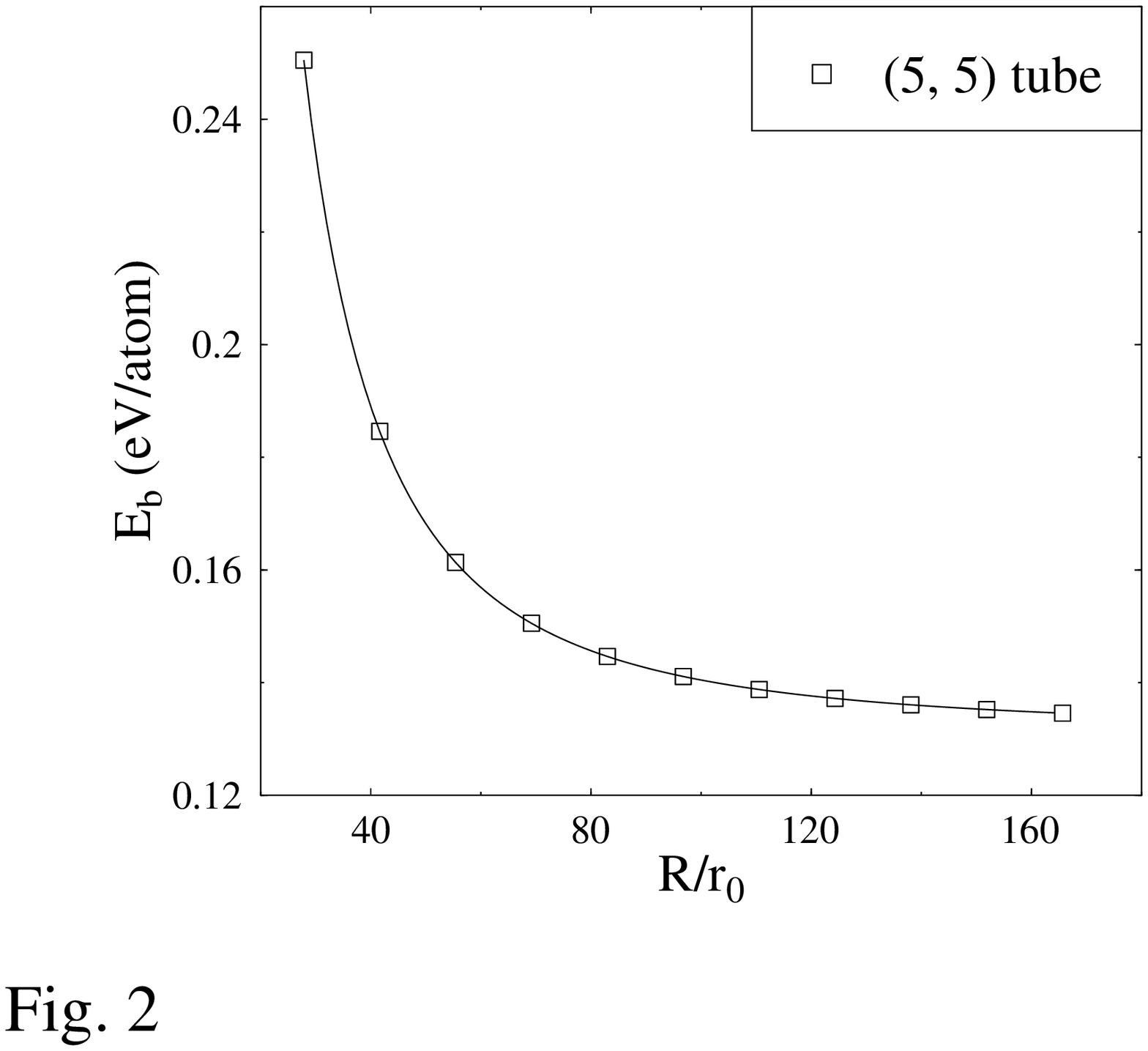}}
\caption{
 Strain energy per atom 
versus the bending radius R in (5,5) tube. 
The solid line is a fit to the $E_r+\lambda/R^2$, where $E_r$
is the straight (5,5) tube strain energy. 
$\lambda=92 \times 1.42^2$ ${\rm eV {\AA}^2/atom}$. 
Here $r_0=1.42$ $\AA$, is the bond length of graphite.
\label{fig2}
}
\end{figure}

\begin{figure}
\centerline{\epsfxsize=10cm \epsfbox{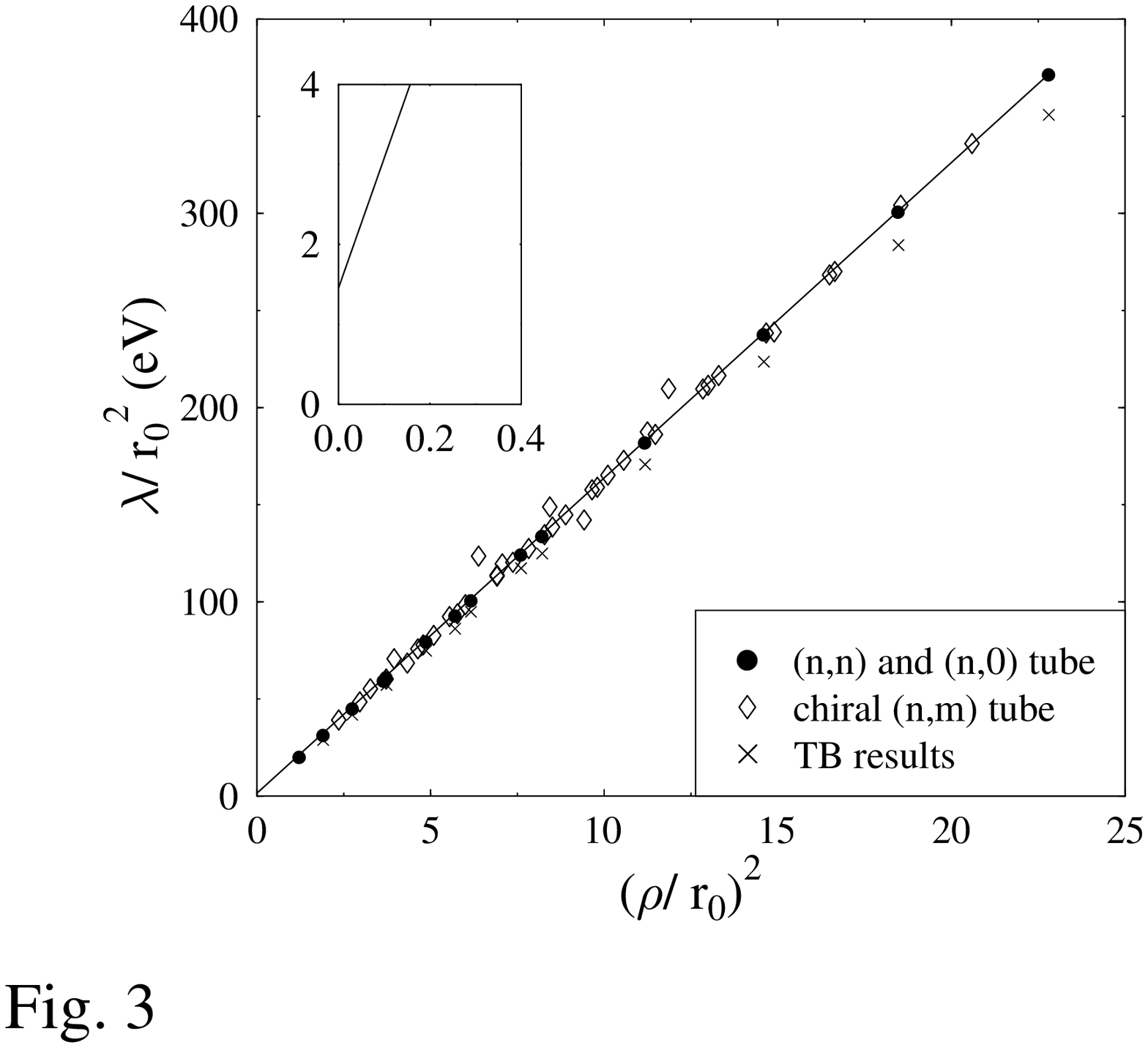}}
\caption{
 the value of $\lambda$ of some (n,0) and (n,n) tubes and some 
 chiral tubes. The solid line
is a fit the all data to $a^{*}+b^{*} {\rho}^2$, 
$a^{*} \approx 1.5 \ {r_0}^2$ ${\rm eV {\AA}^2/atom}$
 $b^{*} \approx 16.2$ ${\rm eV}$. (TB results: $a^{*} \approx 1.05$,
 $b^{*} \approx 15.3$ ${\rm eV}$ in Ref. {\protect\cite{Zhou1}}). 
The inset shows the fitting solid line from all data nearby $\rho=0$.
To compare, we give the TB results, too.
\label{fig3}
}
\end{figure}

\begin{figure}
\centerline{\epsfxsize=10cm \epsfbox{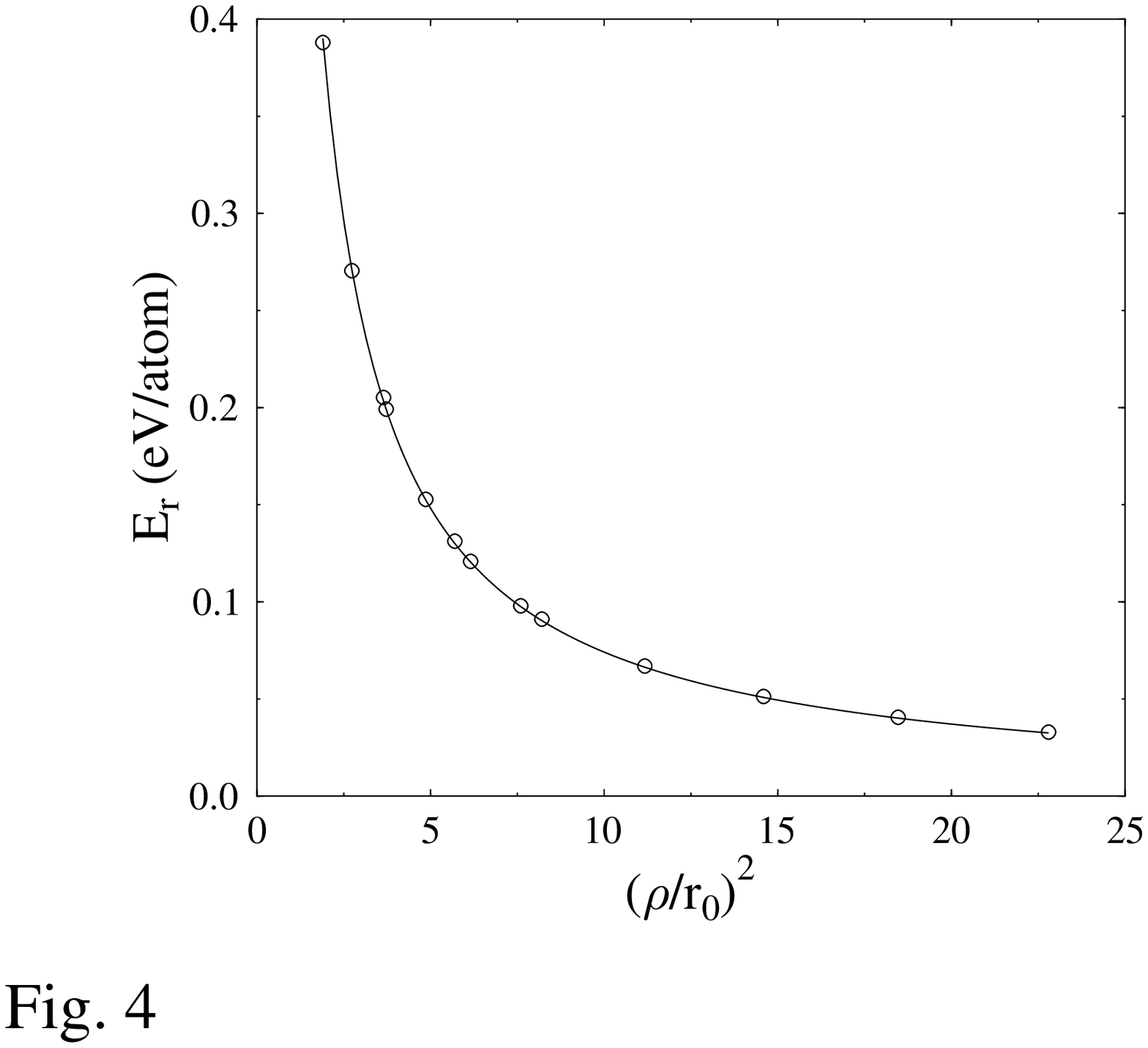}}
\caption{
The strain energy per atom versus the radius of 
$(n, n)$ and $(n, 0)$ tubes. 
The solid 
line drawn across the data corresponds to a least square fit to 
the $\frac{C}{\rho^2}$ behavior. 
${\cal C}=0.74 \cdot {r_0}^2 \approx 1.48$ ${\rm eV {\AA}^2/atom}$. 
${\cal C}$ is independent on the helicity of tubes.  
\label{fig4}
}
\end{figure}

\end{document}